\documentstyle[12pt]{article}

\newcommand{\be}{\begin{equation}}
\newcommand{\ee}{\end{equation}}
\newcommand{\bea}{\begin{eqnarray}}
\newcommand{\eea}{\end{eqnarray}}

\begin{document}

\title{\bf Naked strong curvature singularities in Szekeres space-times}

\author{{\em Pankaj S. Joshi}
\\{Theoretical Astrophysics Group,}
\\{Tata Institute of Fundamental Research,}
\\{Homi Bhabha Road, Bombay 400 005, India}
\medskip
\\{\em Andrzej Kr\'olak}
\\{Institute of Mathematics, Polish Academy of Sciences,}
\\{\'Sniadeckich 8, 00-950 Warsaw, Poland.}}

\date{}
\maketitle
 
\begin{abstract}
We investigate the occurrence and nature of naked singularities in
the Szekeres space-times. These space-times represent irrotational dust.
They do not have any Killing vectors and they are generalisations 
of the Tolman-Bondi-Lema\^{\i}tre space-times.
It is shown that in these space-times there exist naked singularities 
that satisfy both the limiting focusing condition
and the strong limiting focusing condition. The implications of this result 
for the cosmic censorship hypothesis are discussed.
\end{abstract}

\newpage

There exist many explicit solutions of the Einstein equations that
exhibit naked singularities. Nevertheless the {\em cosmic censorship
hypothesis} can still be true if these singularities could be shown
to be in some sense non-generic. It was supposed by Tipler \cite{T}
and independently by Kr\'olak \cite{KMSc} that all generic singularities
are of {\em strong curvature} type. We say that a future-incomplete 
(respectively past-incomplete) causal
geodesic terminates in a strong curvature singularity in the future
(respectively in the past) if for every point $q \in \lambda$ 
the expansion $\theta$ of the congruence
of the future-directed (respectively past-directed) causal geodesics 
originating from q and infinitesimally
neighbouring $\lambda$ diverges ({\bf Tipler's strong curvature singularity}) 
(becomes negative ({\bf Kr\'olak's strong curvature singularity})). 
We say that {\bf strong curvature condition} holds if all future and past
incomplete null geodesics generating an achronal set terminate in a strong
curvature singularity.
It was conjectured by Kr\'olak \cite{CT} that under strong curvature 
condition
cosmic censorship holds. Subsequent work has shown that Kr\'olak's conjecture
is not true. On the one hand the censorship theorems proved
by Kr\'olak \cite{KT} and later by Beem and Kr\'olak \cite{BK} needed
an extra restrictive assumption on causal structure of space-times and
on the other hand explicit examples of naked strong curvature singularities
in Kr\'olak's sense (\cite{ES,C,N}) and in Tipler's sense 
(\cite{JD,DJ1,DJ2}) 
were demonstrated in the Tolman-Bondi-Lema\^{\i}tre (TBL) space-times
representing sphericaly symmetric inhomogeneous collapse of dust and
in Vaidya radiation collapse.
The TBL space-times are special in two senses: they are spherically symmetric
and they have matter in the form of irrotational pressureless dust.
It is interesting to know whether naked strong curvature singularities
occur in more general space-times.
In this letter we show that naked strong curvature singularities occur
in Szekers space-time that do not have any Killing vector. This result
shows that naked strong curvature singularites do not arise as a result
of spherical symmetry. Nevertheless the Szekeres space-times have
the same special form of matter as TBL space-times i.e. irrotational
pressureless dust.

The Szekeres space-time \cite{Sz} is a solution of
Einstein's equations representing irrotational dust
\be
G_{ab} = T_{ab} = \rho u_a u_b, \hspace{3mm} u_a u^a = 1,
\ee
where units are chosen so that $c = 8\pi G = 1$. 
The metric has the diagonal form
\begin{equation}
ds^2 = dt^2 - X^2 dr^2 - Y^2 (dx^2 + dy^2),
\end{equation}
where $(r,x,y)$ are comoving spatial coordinates.
The solution is given by (we consider the case $Y' = 
\frac{\partial Y}{\partial r} \neq 0$)
\begin{equation}
Y = \frac{R(t,r)}{P(r,x,y)}, \hspace{5mm} 
X = \frac{P(r,x,y) Y'(t,r)}{\sqrt{1 + f(r)}},
\end{equation}
where $f(r) > - 1$ and
\begin{eqnarray}
P &=& a(r)(x^2 + y^2) + 2b_1(r)x + 2b_2(r)y + c(r),\\
& & ac - b_1^2 - b_2^2 = \frac{1}{4},\\
& & \dot{R}^2 = f + \frac{F(r)}{R},
\end{eqnarray}
where $F(r)$ is an arbitrary function of $r$ and where dot denotes partial
derivative w.r.t time coordinate $t$.

We assume the following regularity conditions.

1. The metric is everywhere $C^1$. Then the function $P$ must be everywhere 
non-zero and its derivative w.r.t $r$ must be continuous and vanishing at
$r = 0$. 

2. The metric is locally Euclidean at $r = 0$. Then it is necessary to set
\be
f(0) = 0.
\ee

3. The function $R_o(r) = R(r, 0)$ is a monotonically increasing function
of $r$ . We can then use the freedom in the choice of the
radial coordinate $r$ to obtain
\be
R_o(r) = r.
\ee

The dust density $\rho$ is given by
\begin{equation}
\rho = \frac{P F' - 3 F P'}{P^2 R^2 Y'}.
\end{equation}
Although for $P > 0$ the surfaces $r=const$, $t=const$ are spheres, 
the solution is not
spherically symmetric because the spheres are not concentric, their
centers are given by $(-a^{-1}b_1, -a^{-1}b_2)$.
Szekeres has also analysed the singularities and their causal structure
in his space-times.
When $R = 0$, the singularity is of the {\em first kind},
and when $Y'= 0$ the singularity is of the {\em second
kind}. 
The singularities of the second kind are familiar shell-crossing singularities
that also occur in TBL space-times \cite{YSM}. Like in the TBL
space-times shell-crossing singularities in Szekeres spaces can also be both
locally and globally naked \cite{Sz}. However they are generally believed 
to be mild and we shall not consider them here. We shall eliminate these
singularies by imposing a regularity condition
\be
Y' > 0.
\ee
Szekeres has also shown that whenever $r > 0$ the shell of dust always
crosses the apparent horizon before collapsing to singularity and therefore for
$r > 0$ the singularity cannot be naked.
Therefore the singularity of the first kind can be naked only when  $r = 0$
which we call the central singularity.
This situation is analogous to the TBL case.
We shall show that like in TBL space-times naked strong curvature singualrities
do occur in Szekeres space-times.
We shall consider the case of gravitational collapse i.e. we shall require
$\dot{R} < 0$. For simplicity we shall only consider the case
of mariginally bound collapse i.e. we set
\be
f(r) = 0.
\ee
Then the function $R(r,t)$ is given by
\be
R = r \left (1 - \frac{3}{2}\sqrt{\frac{F}{r^3}}t\right )^{2/3}.
\ee
Our analysis follows that of Joshi and Dwivedi for the TBL case \cite{JD}.
We introduce a set of new functions:
\bea
X &=& \frac{R}{r^{\alpha}},\\
\eta &=& r\frac{F'}{F},\\
\Lambda &=& \frac{F}{r^{\alpha}},\\
\Theta &=& \frac{1 - \frac{1}{3}\eta}{r^{\frac{3(\alpha - 1)}{2}}},\\
{\cal L} &=& r \frac{P'}{P},
\eea
where $\alpha \geq 1$ and the unique value of the constant $\alpha$
is determined by the condition that $\frac{\Theta}{\sqrt{X}}$ does not vanish 
or goes to infinity identically as $r \rightarrow 0$ in the limit of
approach to the central singularity along any $X = const$ direction.
We shall assume that the above functions are at least $C^2$.
Partial derivatives $R'$ and $\dot{R}'$ that are important in the analysis
of the singularity are then given by
\bea
\dot{R} &=& - \sqrt{\frac{\Lambda}{X}},\\
R' &=& r^{\alpha - 1} H,\\
\dot{R}' &=& - \frac{N}{r},
\eea
where
\bea
H &=& \frac{1}{3} \eta X + \frac{\Theta}{\sqrt{X}},\\
N &=& - \frac{\sqrt{\Lambda}}{2 X^2}\left (\Theta - 
\frac{2}{3}\eta X^{3/2}\right ).
\eea
The tangents $K^a = dx^a/dk$ for the outgoing radial ($x = const, y = const$)
null geodesics can be written as
\bea
K^t &=& \frac{dt}{dk} = \frac{{\cal P}}{Y},\\
K^r &=& \frac{dr}{dk} = \frac{{\cal P}}{P Y Y'},\\
K^x &=& \frac{dx}{dt} = 0,\\
K^y &=& \frac{dy}{dt} = 0,
\eea
where ${\cal P}$ satisfies the differential equation
\be
\frac{d{\cal P}}{dk} + {\cal P}^2\left (\frac{\dot{Y}'}{Y Y'} - 
\frac{\dot{Y}}{Y^2} - \frac{1}{P Y^2}\right ) = 0. \label{G}
\ee
The parameter $k$ is an affine parameter along the null geodesics.

If the outgoing null geodesics are to terminate in the past at the central
singularity $r = 0$, which occurs at some time $t = t_o$ at which
$R(t_o,0) = 0$, then along such geodesics we have $R \rightarrow 0$
as $r \rightarrow 0$. The following is satisfied along null geodesics
\bea
\frac{dR}{du} &=& \frac{1}{\alpha r^{\alpha - 1}}[\dot{R}\frac{dt}{dr} + R'] =\\
& & \left (1 - \sqrt{\frac{\Lambda}{X}}\right )\frac{H(X,u)}{\alpha} -
\frac{\sqrt{X\Lambda}}{\alpha} {\cal L} \equiv U(X,u),
\eea
where we have put $u = r^{\alpha}$.
Let us consider the limit $X_o$ of the function $X$ along the null
geodesic terminating at the sigularity at $R = 0, u = 0$.
Using the l'Hospital rule we get
\bea
X_o = \lim_{R \rightarrow 0, u \rightarrow 0} \frac{R}{u} =
\lim_{R \rightarrow 0, u \rightarrow 0} \frac{dR}{du} =
\lim_{R \rightarrow 0, u \rightarrow 0} U(X,u) = U(X_o,0)
\eea
The necessary condition for the existence of the null geodesic
outgoing from the central singularity is the existence of the
positive real root $X_o$ of the equation
\be
V(X) \equiv U(X,0) - X = 0.
\ee
By our regularity conditions we have that 
$\lim_{r \rightarrow 0} {\cal L} = 0$. Consequently the
necessary condition for the existence of the naked singularity
in the mariginally bound case of Szekeres space-time is the existence of 
positive real root of the equation
\be
\left (1 + \sqrt{\frac{\Lambda_o}{X}}\right )\frac{H(X,0)}{\alpha} - X = 0,
\label{V}
\ee
where we put
\bea
\eta_o = \eta(0),\\
\Lambda_o = \Lambda(0),\\
\Theta_o = \Theta(0).
\eea
This is exactly the same equation as in the mariginally bound TBL case 
\cite{JD}. Consequently the same analysis as in the TBL case applies here.
We shall only summarize here a few key results. To show that the singularity
is naked we still need to prove that there exists a solution of the geodesic
equation such that the tangent $X_o$ is realized at the singularity.
One can prove that there is always at least a single null geodesic 
outgoing from the central singularity (\cite{JD} p.5363).
{\em Thus the existence of the real and positive root of the equation (\ref{V})
is both necessary and sufficient condition 
for the existence of a naked singularity.}

We shall next investigate the strength of the singularity. 
Let $M$ denote the space-time manifold.
Let $J(k)$ be defined on a null geodesic 
$\lambda : (k_o,0] \rightarrow M$ parameterised 
by an affine parameter $k$
\be
J(k) = \int^{0}_{k_o} R_{ab}K^aK^b dk'
\ee
We say that the {\em limiting focusing condition} (LFC) holds if $J(k)$
is unbounded in the interval $(k_o,0]$ and we say that the {\em strong
limiting focusing condition} (strong LFC) if $J(k)$ is non-integrable
on an interval $(k_o,0]$. It is proved in Ref.\cite{CK} that 
LFC implies that $\lambda$ terminates in Kr\'olak's 
strong curvature singularity in the future whereas strong LFC implies 
that $\lambda$ terminates in Tipler's 
strong curvature singularity in the future.

To find out whether the naked singularity satisfies 
strong limiting focusing condition we shall investigate the limit
$\lim_{k \rightarrow 0}k^2 R_{ab}K^aK^b$ along the future directed null
geodesics coming out from the singularity. 
Using the l'Hospital  rule, regularity conditions and Eq.\ref{G} we have
\bea
\lim_{k \rightarrow 0}k^2 R_{ab}K^aK^b = 
\lim_{k \rightarrow 0}\frac{k^2 (F' - 3 F\frac{P'}{P})(K^t)^2}
{R^2 (R' - R\frac{P'}{P})} =
%\lim_{k \rightarrow 0}\frac{\eta\Lambda}{H X^2}
%\left (\frac{k}{u Y \frac{1}{{\cal P}}}\right )^2 = 
\frac{\eta_o \Lambda_o H_o}
{X_o^2 (\alpha - N_o)^2},
\eea
where $N_o = N(X_o,0)$.
Thus if $\Lambda_o \neq 0$ the naked singularity satisfies the
strong limiting focusing condition.

Suppose then that $\Lambda_o = 0$. We can write the Eq.\ref{G}
in the following form
\be
\frac{d (\ln K^t)}{d k} = \left(N - \sqrt{\frac{\Lambda}{X}} {\cal L}\right )
\frac{1}{r}\frac{dr}{dk}.
\ee
Since as $k \rightarrow 0$, $N \rightarrow 0$ because $\Lambda_o = 0$
and ${\cal L} \rightarrow 0$ by regularity condition the right hand
side of the above equation is integrable on the interval $(k_o,0]$ with
respect to $k$. Thus the limit $\lim_{k \rightarrow 0} K^t$ exists.
We also have
\be
K^r = \frac{K^t}{r^{\alpha - 1}(H - X {\cal L})}.
\ee
Thus if $\alpha = 1$ the limit $\lim_{k \rightarrow 0} K^r$ also exists. 
Let us suppose thet $\alpha = 1$ and let us consider the limit $\lim_{k \rightarrow 0}k R_{ab}K^aK^b$
Applying the l'Hospital rule twice and using Eq.\ref{G} we get
\bea
\lim_{k \rightarrow 0}k R_{ab}K^aK^b =
\frac{\eta_o  \Lambda_o' H_o}{X_o^2 (\alpha - N_o)^2} 
\lim_{k \rightarrow 0}K^r.
\eea
Thus when $\alpha = 1$ the above limit is finite and the naked singularity
just satisfies the limiting focusing condition ($R_{ab}K^aK^b$ diverges
logarithmically).
If $\alpha > 1$ the above limit diverges and the naked singularity also
satisfies the limiting focusing condition (though not strong focusing 
condition unless $\Lambda_o \neq 0$).
The above results show that under the regularity conditions given above
and for mariginally bound collapse any central naked singularity
in the Szekeres space-time is always a strong curvature singularity 
in Kr\'olak's
sense.

\section{Acknowledgements}
This work was supported in part by Polish Science Council 
grant KBN 2 P301 050 07.

\end{document}